\documentclass[aps,prd,prabib,twocolumn,showpacs,nofootinbib]{revtex4}
\usepackage{graphicx} \usepackage{amsmath} \usepackage{amssymb}
\usepackage{amsfonts} \usepackage{bm}

\begin{document}

\newcommand{\be}{\begin{equation}} \newcommand{\ee}{\end{equation}}
\newcommand{\bea}{\begin{eqnarray}}\newcommand{\eea}{\end{eqnarray}}


\title{Quantization of exciton in magnetic field background}

\author{Pulak Ranjan Giri} \email{pulakranjan.giri@saha.ac.in}

\affiliation{Theory Division, Saha Institute of Nuclear Physics,
1/AF Bidhannagar, Calcutta 700064, India}

\author{S. K. Chakrabarti} \email{sayan.chakrabarti@saha.ac.in}

\affiliation{Theory Division, Saha Institute of Nuclear Physics,
1/AF Bidhannagar, Calcutta 700064, India}

\begin{abstract}
  The possible mismatch between the theoretical and
  experimental absorption of the edge peaks in semiconductors in a magnetic
  field background may arise due to the approximation scheme used to
  analytically  calculate the absorption coefficient.
  As a possible
  remedy we suggest to consider nontrivial boundary conditions on
  $x$-$y$ plane by in-equivalently quantizing the exciton in
  background magnetic field. This inequivalent quantization is based
  on von Neumann's method of self-adjoint extension, which is
  characterized by a parameter $\Sigma$. We obtain bound state solution and
  scattering state solution, which in general depend upon the self-adjoint
  extension parameter $\Sigma$. The  parameter $\Sigma$ can be used to fine
  tune the optical absorption coefficient  $K(\Sigma)$  to match  with the
  experiment.

\end{abstract}



\date{\today}

\maketitle

\section{Introduction}

Exciton \cite{frenkel} is a bound pair between an electron and  a hole in a
crystal, which interact through Coulomb attraction between them.
When an electron from the valence band is excited into the conduction
band, the missing electron in the valence band leaves a hole 
with opposite electric charge. The exciton results from the binding of
the electron with its hole. As a result the exciton has slightly less
energy than the unbound electron and hole. It has an important role to
play in the fine structure of the absorption edge in semiconductor. It
has been seen that as the photon energy goes above the band gap energy
of the semiconductor, the absorption coefficient shows spikes due to
the formation of exciton \cite{elliott}. This is an important study no
doubt, because it explores  the band structure and the character of
the one electron states of the semiconductor. The properties of a
semiconductor are mostly determined by the electronic states which lie
just below and just above the forbidden gap.

An appreciable amount of studies have been made to explore the band
structure of semiconductors by examining direct and indirect optical
absorption in  semiconductors \cite{hall}. This absorption may arise
due to the transition between the states of the same side of the
forbidden gap or across the forbidden gap.

There have been a lot of studies in this field both theoretically and
experimentally, see for example \cite{altarelli,rom}. Since the
constituents of exciton are electrically charged, the magnetic fields
have certain effects on the energy spectra of the exciton and it in
turn affects the fine structure of the absorption.  The magnetic fields
have a Zeeman like effect on the energy spectra of an exciton, which has
been investigated by studying optical absorption in many substances,
for example, in $InSb$ \cite{burstein}, $Ge$ and $InAs$. Magneto-optic
effect in strained and unstrained $Ge$ have been performed in
\cite{ge}. In \cite{elliott1}, a theoretical study of the effect of
magnetic field on the absorption edge in solid has been done,
neglecting the Coulomb interaction between the electron and the hole. It
is shown that a series of peaks may form due to the transition
between the sub-bands, which originate due to the splitting of
electron energy states in the solid by the magnetic field. Instead of
forming peaks, in some cases a series of steps are formed due to
magnetic field. But Coulomb interaction certainly has effect
\cite{dresselhaus} on the absorption in solids. It is therefore
required to consider Coulomb interaction and magnetic field
simultaneously in the study of absorption \cite{elliott2}. In general
this problem is hard to tackle analytically for any magnetic field.
But for large magnetic field an approximation scheme \cite{elliott2}
can be used to handle the situation and can be checked experimentally
for some substances like $InSb$, $Ge$ etc. According to this
approximation scheme, which is valid in large magnetic field, the
Coulomb term should only affect the dynamics of the exciton in the
direction of the applied magnetic field and it is therefore possible
to reduce the time independent three dimensional Schr\"{o}dinger
equation into one dimensional eigenvalue equation in the direction of
the magnetic field. The inclusion of the Coulomb interaction shows
that the the absorption peaks in each magnetic sub-band transition is
moved towards the lowest exciton line. The other effects are the
reduction of intensity of the continuous absorption and to smooth out
the peaks near the absorption edge.  The peaks are the exciton peaks
instead of the magneto-optic peaks.

It is  expected that the analytical theory of absorption in solids
\cite{elliott2} will not be able to exactly explore the true nature
of the band structure due to the approximation used in the
calculation. In this article, we therefore try to incorporate
some short range interactions into the boundary condition, imposed
on the domain of the system so that the results based on the
approximate model agrees with the experiments as much as possible.
This nontrivial boundary condition will in general change the energy
spectrum of the system and it will depend on a parameter, called
self-adjoint extension parameter.  The fine structure  of the
optical absorption edge in a crystal due to exciton essentially
depends on the energy spectrum of the exciton. It is therefore
expected to change if the spectrum changes. To get a consistent
boundary condition  and the domain, so that the Hamiltonian remains
self-adjoint, we  perform a self-adjoint extension (SAE) of the one
dimensional Hamiltonian in the direction of the magnetic field by
using von Neumann's method \cite{reed}. Self-adjointness of a
Hamiltonian is essential, because otherwise the Hamiltonian would
generate complex eigenvalues and the time evolution of the states
will not be unitary. SAE has received lot of interests in recent
years  and is now being used extensively in different branches of
physics, to explore the nontrivial quantum behavior of different
systems \cite{kumar1,kumar2,kumar3,biru,stjep,giri,bh}.


The present article has been  organized as follows: In sec. II, we
discuss about the exciton in a magnetic field background and write
an effective radial eigenvalue equation following \cite{elliott2},
which we need to quantize using nontrivial boundary condition. The
method of self-adjoint extensions  (SAE) is discussed in Sec. III,
which is essential to get nontrivial boundary condition for the
effective Hamiltonian $H_{|z|}$ so that the Hamiltonian is
self-adjoint. The bound state and scattering state solutions are
discussed in Sec. IV and V respectively. We conclude the paper with
a discussion in Sec VI.

\section{Exciton in magnetic field background}

We consider dynamics of electron and hole pair with effective masses
$m_e$ and $m_h$ respectively in a medium of dielectric constant
$\kappa$. In a magnetic field background $\bf B$ the time independent
Schr\"{o}dinger
equation for the system is given by \cite{elliott2}
\begin{eqnarray}
\nonumber\left[\frac{1}{2m_e}\left(\boldsymbol{p}_e+\frac{e}{c}\boldsymbol{A}(\boldsymbol{r}_e)\right)^2
  + \frac{1}{2m_h}\left(\boldsymbol{p}_h-\frac{e}{c}\boldsymbol{A}(\boldsymbol{r}_h)\right)^2
  \right.\\ \left. -
\frac{e^2}{\kappa|\boldsymbol{r}_e- \boldsymbol{r}_h|}\right]\Psi= E\Psi\,,
\end{eqnarray}
where $\boldsymbol{A}(\boldsymbol{r}_e)$ and $\boldsymbol{A}(\boldsymbol{r}_h)$ are magnetic vector
potentials at electron and hole positions respectively.

Considering the fact that the states of interest in optical
transitions are only those in which
$\boldsymbol{k}=\boldsymbol{k}_e+\boldsymbol{k}_h=0$ (where
$\boldsymbol{k}_e$ and $\boldsymbol{k}_h$ are electron and hole wave
vectors respectively), Lamb \cite{lamb} simplified the above equation
by changing the variables to the center of mass coordinate
$\boldsymbol{p}=m_e\boldsymbol{r}_e+m_h\boldsymbol{r}_h/m_e+m_h$ and
relative position $\boldsymbol{r}=\boldsymbol{r}_e-\boldsymbol{r}_h$
and using the substitution $\Psi(\boldsymbol{r}_e,
\boldsymbol{r}_h)=U(\boldsymbol{r})\exp[-\frac{ie}{2\hbar
  c}\boldsymbol{B}\times \boldsymbol{p}.\boldsymbol{r}]$

\begin{eqnarray}
\left[-\frac{\hbar^2}{2\mu}\nabla^2-\frac{ie\hbar}{2c}\left(\frac{1}{m_e}
-\frac{1}{m_h}\right)\boldsymbol{B}.\boldsymbol{r}\times\boldsymbol{\nabla}+
\right.\nonumber\\
\left. \frac{e^2}{8\mu
c^2}(\boldsymbol{B}\times\boldsymbol{r})^2-\frac{e^2}{\kappa r}
\right]U(\boldsymbol{r})= EU(\boldsymbol{r})\,,\label{Ham}
\end{eqnarray}
where $\mu$ is the reduced  mass $\mu=m_em_h/(m_e+m_h)$ of the
electron-hole system. The presence of Coulomb interaction  spoils
the separability of (\ref{Ham}) in any co-ordinates. So it is
difficult to solve the equation in any coordinate system. It can
however be reduced in an one dimensional eigenvalue equation in
large magnetic field limit. Note that the magnetic field affects the
motion of a charged particle in a direction perpendicular to the
magnetic field. In our case we take the magnetic field in $z$
direction. So, in $z$ direction the charged particle motion will be
a free particle motion in absence of Coulomb force. But in presence of Coulomb
force the free motion in $z$ direction will be modified according to Ref.
\cite{schiff}. The potential along $z$ direction will now be the average
Coulomb potential over the $x$-$y$ plane at each point in $z$ axis 
\cite{elliott2}.
From now on, we take $\hbar^2=2\mu=e^2=1$, for the simplicity of
calculations. In these units, the one dimensional Differential
Equation is of the form \cite{elliott2}
\begin{eqnarray}
\nonumber H_{|z|}\chi(|z|)&\equiv& -\frac{d^2\chi}{d|z|^2}-
\left[\frac{1}{\kappa(a+|z|)}
-\frac{Aa}{\kappa(a+|z|)^2}\right]\chi(|z|)\\
& =& E_{|z|}\chi(|z|)\,, \label{1dhamiltonian}
\end{eqnarray}
where $a$ and $A$ are constant parameters (see Ref. \cite{elliott2} for
details) and its values will depend on the  eigenfunctions 
on the perpendicular 
$x$-$y$ plane. We thus lead to a one dimensional differential eigenvalue
problem which depends on the modulus of  Cartesian co-ordinate $z$ due to the
reflection symmetry on the $z$ axis.  The inner product is now defined as
\begin{eqnarray}
\left(\chi_1(|z|),\chi_2(|z|)\right)=\int_0^\infty\chi_1(|z|)\chi_2(|z|)d|z|
\end{eqnarray}
where $d|z|$ is the measure.
In order to find the bound state eigenvalue and corresponding
eigenfunction of (\ref{1dhamiltonian}), we need to consider a
physically meaningful boundary condition. We want to find out this
boundary condition by von Neumann's method of self-adjoint extension (SAE).

\section{von Neumann's method of SAE}

Before going into the actual Hamiltonian $H_{|z|}$ of
(\ref{1dhamiltonian}), it is essential to discuss about the general
properties and method of self-adjoint extensions of an operator
$\mathcal B$ defined over the elements of a Hilbert space $\mathcal
H$. Let us consider $\mathcal{D}(\mathcal B)$ as the domain of
$\mathcal B$ such that the operator $\mathcal B$ becomes symmetric in
that domain. The operator $\mathcal B$ is said to be symmetric (or
hermitian) if it satisfies the condition
$\left(\mathcal{B}\phi_1,\phi_2\right)=
\left(\phi_1,\mathcal{B}\phi_2\right)$, $\forall \phi_1,\phi_2
\in\mathcal{D}(\mathcal B)$, where $(.,.)$ is the inner product
defined over the Hilbert space. Given the operator $\mathcal B$ and
the inner product $(.,.)$, one can calculate the adjoint operator
$\mathcal {B}^*$ from the Green's formula \cite{dunford}. This adjoint
operator $\mathcal {B}^*$ should have a domain $\mathcal{D}(\mathcal
{B}^*)$, which can be obtained from
$\left(\mathcal{B}^*\phi_1,\phi_2\right)=
\left(\phi_1,\mathcal{B}\phi_2\right)$, $\forall \phi_2
\in\mathcal{D}(\mathcal B)$, where $\phi_1\in \mathcal{D}(\mathcal
{B}^*)$. Now, if the operator is self-adjoint then
$\mathcal{D}(\mathcal {B})=\mathcal{D}(\mathcal {B}^*)$ and for non
self-adjoint case $\mathcal{D}(\mathcal {B})\neq \mathcal{D}(\mathcal
{B}^*)$. For non self-adjoint operators we need to find out
self-adjoint extensions, because according to Stone's theorem
\cite{reed} self-adjointness of an operator is necessary in order to
guarantee unitary evolution. We follow von Neumann's method for
finding self-adjoint extensions of the Hamiltonian, if there is any.
According to this method, we first need to find out the solutions of
the equation $\mathcal {B}^*\phi^\pm=\pm i\phi^\pm$. $\phi^\pm$ are
called deficiency space solutions, because the numbers $n^\pm$ are a
measure by which a non self-adjoint operator is away from
self-adjointness, where $n^\pm$ denotes the number of $\phi^\pm$
solutions. For $n^+=n^-=0$, the operator $\mathcal {B}^*$ is
self-adjoint. For $n^+=n^-=n\neq 0$, the operator $\mathcal {B}^*$ is
not self-adjoint, but admits self-adjoint extensions characterized by
$n^2$ parameters of $U(n)$. Finally, for $n^+\neq n^-$, the operator
$\mathcal {B}^*$ is not self-adjoint and can't be made self-adjoint.
The self-adjoint extension of the operator $\mathcal {B}^*$ is given
by the domain $\mathcal{D}_{U(n)}(\mathcal {B})\equiv \phi+ \phi^+ +
U(n)\phi^-$, where $\phi\in \mathcal{D}(\mathcal {B})$.
\begin{figure}
\includegraphics[width=0.45\textwidth, height=0.18\textheight]{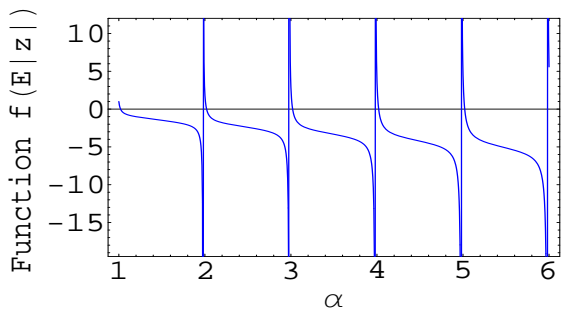}
\caption {(color online) A plot of the function of the eigenvalue $f(E_{|z|})$
as a function of the parameter $\alpha$. This type of plot is previously
obtained in Ref. \cite{kumar2,feher}. This discussion and the properties of
the function can be found in \cite{feher}.}
\end{figure}

\section{Bound state solution of exciton}

We now come to the discussion of the Hamiltonian
(\ref{1dhamiltonian}), for which we seek a self-adjoint domain. We
construct an initial domain 
$\phi(|z|)\in\mathcal{D}(H_{|z|})$, where $\phi(|z|)$ and  $\phi'(|z|)$ are
absolutely continuous and $\phi(0)= \phi'(0)=0$. It is implied that
$\phi(|z|)$ and $H_{|z|}\phi(|z|)$ are square-integrable functions, i.e.,
belong to the Hilbert space. Note that $\mathcal{D}(H_{|z|})$ is
so restricted that the
Hamiltonian $H_{|z|}$ becomes  symmetric in this domain. The adjoint
Hamiltonian $H^*_{|z|}$ has the same differential form
(\ref{1dhamiltonian}), but the domain is given by
$\phi(|z|)\in\mathcal{D}(H^*_{|z|})$, where  $\phi(|z|)$ and  $\phi'(|z|)$ are
absolutely continuous.
The two domains are seen
to be different, i.e., $\mathcal{D}(H_{|z|})\neq
\mathcal{D}(H^*_{|z|})$. So we need to find out the deficiency space
solutions (square integrable), which are in our case the solutions
of the differential equations $H^*_{|z|}\phi_\pm = \pm i\phi_\pm$.
These are
\begin{eqnarray}
\phi_\pm(|z|)=
W_{\alpha_\pm,m}(\frac{1}{\kappa\alpha_\pm}(a+|z|))\,,\label{def}
\end{eqnarray}
where $\alpha_\pm=\pm\frac{1}{2\kappa}ie^{\mp i\pi/4}$, $m=
\sqrt{(\frac{1}{4}+\frac{Aa}{\kappa})}$ and $W$ is the Whittaker's
function \cite{abr}. The square integrability of (\ref{def}) can be
easily checked from the asymptotic and short distance behavior. 
Note however that the potential at $|z|\to 0$ is always finite so there is no
problem of short distance square-integrability but the short distance behavior
will help to get the spectrum for the problem.
The asymptotic behavior is given by
\begin{eqnarray}
\lim_{|z|\to\infty}\phi_\pm(|z|)\simeq{(\frac{1}{\kappa\alpha_\pm}(a+|z|))}^{\alpha_\pm}
\exp(-\frac{1}{2\kappa\alpha_\pm}(a+|z|))\,, \label{adef}
\end{eqnarray}
which goes to zero asymptotically, i.e.,  $\phi_\pm(|z|)\to 0$. In order to
get short distance behavior we need to express the Whittaker's function in
terms of Kummer function as
\begin{eqnarray}
\nonumber W_{a,b}(x) =e^{-\frac{1}{2}x}x^{\frac{1}{2}+b}\frac{\pi}
{\sin\pi(1+2b)}\times~~~~~~~~~~~~~~~~~~~~~~\\
\left[\frac{M(\frac{1}{2}+b-a,1+2b,x)}{\Gamma(\frac{1}{2}-a-b)\Gamma(1+2b)}-
x^{-2b}\frac{M(\frac{1}{2}-a-b, 1-2b,
  x)}{\Gamma(\frac{1}{2}+b-a)\Gamma(1-2b)}\right]\,,
\label{}
\end{eqnarray}
where $M$ is the Kummer function \cite{abr}. The short distance
behavior can now be evaluated as
\begin{eqnarray}
\nonumber\lim_{|z|\to
0}\phi_\pm(|z|)\simeq~~~~~~~~~~~~~~~~~~~~~~~~~~~~~~~~~~~~~~~~~~
\\\left[\mathcal{A}_\pm.(a+|z|)^{1/2+m}-
\mathcal{B}_\pm.(a+|z|)^{1/2-m}\right]\,,\label{sdef}
\end{eqnarray}
where $\mathcal{A}_\pm=
\frac{\pi}{\sin\pi(1+2m)\Gamma(1/2-\alpha_\pm-m)\Gamma(1+2m)}$ and
$\mathcal{B}\pm=
\frac{\pi}{\sin\pi(1+2m)\Gamma(1/2-\alpha_\pm+ m)\Gamma(1-2m)}$.

Since in our case $n_+=n_-=1$, $H_{|z|}$ is not self-adjoint but it
admits self-adjoint extensions, characterized by a parameter
$\Sigma$ and the domain over which the Hamiltonian $H^\Sigma_{|z|}$ is
self-adjoint is given by
\begin{eqnarray}
\mathcal{D}_{\Sigma}(H^\Sigma_{|z|})= \{\phi(|z|) +\phi_+(|z|) +
\exp(i\Sigma)\phi_-(|z|) \}\,, \label{sdomain}
\end{eqnarray}
where $\phi(|z|)\in\mathcal{D}(H_{|z|})$.
Now we can solve the eigenvalue problem (\ref{1dhamiltonian}) and find out
eigenvalue using the extended  domain (\ref{sdomain}). The eigenfunction of
(\ref{1dhamiltonian}) apart from normalization is given by
\begin{eqnarray}
\chi(|z|)= W_{\alpha, m}(\frac{1}{\kappa\alpha}(a+|z|))\,,
\label{eigenfunction}
\end{eqnarray}
where $\alpha = \frac{1}{2\kappa}\sqrt{-\frac{1}{E_{|z|}}}$ and $m$ has
already been defined.
The eigenvalue of (\ref{eigenfunction}) can be found by looking at the short
distance behavior of (\ref{sdomain}) and  (\ref{eigenfunction}) and equating
the coefficients of equal powers of $(a+|z|)$. For general value of the
self-adjoint extension parameter $\Sigma$, the eigenvalue equation is
found to be
\begin{eqnarray}
f(E_{|z|})\equiv \frac{\Gamma(1/2+m-\alpha)}{\Gamma(1/2-m-\alpha)}=
\frac{\chi_1\cos(\theta_1-\Sigma/2)}{\chi_2\cos(\theta_2-\Sigma/2)}\,,
\label{eigenvalue}
\end{eqnarray}
where $\Gamma(1/2-m-\alpha_+)=\chi_1\exp(-i\theta_1)$ and
 $\Gamma(1/2+ m-\alpha_+)=\chi_2\exp(-i\theta_2)$. The eigenvalue equation
(\ref{eigenvalue}) is a 1-parameter family of equation. It is to be noted
 that the equation of the form (\ref{eigenvalue}) is discussed in detail in
Ref. \cite{kumar2,feher} and the reason we get similar equation here is that
 the nature of the boundary condition obtained from the short distance
 behavior of the special functions are similar in both cases. Each value
of the parameter $\Sigma$ correspond to a different quantization
for the system. A plot of $f(E_{|z|})$ as a
function of $\alpha$ can be found in FIG. 1. The intersection of the
horizontal line with the curves between consecutive vertical lines
gives the value of alpha, in other words value of the energy
$E_{|z|}$. It is in general hard to solve (\ref{eigenvalue})
analytically. But in certain limits it can be evaluated exactly. For
example, when the R.H.S of (\ref{eigenvalue}) is zero, the
eigenvalue is
\begin{eqnarray}
E_{|z|}=- \frac{1}{4\kappa^2(1/2-m+n)^2}\,, \label{eigen1}
\end{eqnarray}
where $n=0, 1, 2, 3,...$. When the R.H.S is infinity, the eigenvalue
$E_{|z|}$ becomes
\begin{eqnarray}
E_{|z|}=- \frac{1}{4\kappa^2(1/2+m+n)^2}\,. \label{eigen2}
\end{eqnarray}
In between zero and infinity the eigenvalue can be obtained from
FIG. 1. Let us now consider the infinite magnetic field limit,
$\boldsymbol{B}\to\infty$. In this limit $m\to 1/2$ and from
(\ref{eigen1})  we get
\begin{equation}
E_{|z|}= -\frac{1}{4\kappa^2n^2}\,,~~~n=0,1,2,... \label{eigen3}
\end{equation}
Note that for $\boldsymbol{B}\to\infty$, we get the same energy
eigenvalue (\ref{eigen3}) from  Eq. (\ref{eigen2}), but this time
$n=1,2,3,..$.

\section{Scattering state solution of exciton}

We now move to the discussion of  scattering state solutions. The eigenvalue
equation for scattering states will be the same Eq. (\ref {1dhamiltonian}),
where $E_{|z|}$ is now positive. The scattering state solution is of the form
\begin{eqnarray}
\nonumber\chi(|z|)= C(i\tilde\alpha)M_{i\tilde\alpha,m}\left(\frac{1}{i\kappa\tilde\alpha}(a+|z|)\right) +\\
D(i\tilde\alpha)W_{i\tilde\alpha,m}\left(\frac{1}{i\kappa\tilde\alpha}(a+|z|)\right)
\label{scattering_state}
\end{eqnarray}
where $\tilde\alpha= \frac{1}{2\kappa}\sqrt{\frac{1}{E_{|z|}}}$ and $
C(i\tilde\alpha)$ and  $D(i\tilde\alpha)$ are undetermined coefficients. One
can now determine the coefficients $C$ and $D$ by looking at the short
distance behavior of (\ref{sdomain}) and (\ref{scattering_state}) and
equating the same powers of $(a+|z|)$, which will now depend upon the
self-adjoint extension parameter $\Sigma$. Note that the scattering state
solution (\ref{scattering_state}) does not belong to the domain  
(\ref{sdomain}) for the simple reason that the scattering state solutions are
not normalizable and thus does not belong to the Hilbert space. But there is
no problem for the scattering states to be square integrable at  short
distance and thus it may belong to the domain (\ref{sdomain}) as far as the
short distance behavior is concerned (see \cite{reed} for detail 
discussion with an example)
It is also possible to calculate
the $S$-matrix and phase shift from this scattering state solution.

\section{Discussion}
The knowledge of quantum mechanical behavior of exciton in a magnetic
field background is essential for the study of absorption in
semiconductors as has been shown in Ref. \cite{elliott2}. Since the time independent Schr\"{o}dinger equation
for the exciton cannot be exactly solved, an approximation scheme is
generally used \cite{elliott2} in order to solve the system
analytically. In order to overcome the shortcomings of the
approximation scheme and to guarantee unitary evolution of the system
we obtain a 1-parameter family of self-adjoint extensions of the
effective one dimensional eigenvalue equation. We obtain bound state
and scattering state solutions, which in general depend on the
self-adjoint extension parameter $\Sigma$. The direct optical
absorption coefficients $K(\Sigma)$ depend upon the bound and
scattering state solutions which are dependent on the boundary
conditions characterized by the parameter $\Sigma$. Therefore the
parameter $\Sigma$ can be used to fine tune the optical absorption
coefficient  $K(\Sigma)$  to match  with the experimental data. Our procedure
gives all possible solutions for the system.

\vskip 0.5 cm



\begin{thebibliography}{99}

\bibitem{frenkel} J. Frenkel, Phys. Rev {\bf 37}, 17 (1931).

\bibitem{elliott} R. J. Elliott, Phys. Rev. {\bf 108}, 1384 (1957).

\bibitem{hall} L. H. Hall, J. Bardeen and F. J. Blatt,  Phys. Rev {\bf 95},
559 (1954);
  
G. G. Macfarlane and V. Roberts, Phys. Rev. {\bf 97}, 1714
(1955); 
 
G. G. Macfarlane and V. Roberts, Phys. Rev. {\bf 98}, 1865 (1955);

W. C. Dash and R. Newman, Phys. Rev. {\bf 99}, 1151 (1955);

H Y Fan,  Rep. Prog. Phys. {\bf 19}, 107 (1956).

\bibitem{altarelli} E. J. Johnson, Phys. Rev. Lett. {\bf 19}, 352
  (1967); 

A. Baldereschi and N. O. Lipari, Physical Review Letters {\bf
    25} 373 (1970); 

M. Altarelli and N. O. Lipari, Phys. Rev.{\bf B7},
  3798 (1973); 

M. Altarelli and N. O. Lipari, Phys. Rev. {\bf B9},
  1733 (1974); 

W. Andreoni, M. Altarelli, and F. Bassani, Phys. Rev.
  {\bf B 11}, 2352 (1975).

\bibitem{rom} J. Maultzsch et. al,  Phys. Rev. {\bf B 72}, 241402(R) (2005);

S. Zaric et. al,  Phys. Rev. Lett. {\bf 96}, 016406 (2006);

S. Uryu and T. Ando,  Phys. Rev. {\bf B 74}, 155411 (2006);

S. Uryu and T. Ando,  Phys. Rev. {\bf B 76}, 115420 (2007).


\bibitem{burstein} E. Burstein and G. S. Picus, Phys. Rev. {\bf
105}, 1123 (1957).

\bibitem{ge} D. F. Edwards and V. J. Lazazzera, Phys. Rev. {\bf 120},
  420 (1960).

\bibitem{elliott1} R. J. Elliott, T.P. Mclean and G.G. Macfarlane,
Proc. Phys. Soc. Lond. {\bf 72}, 553 (1958).




\bibitem{dresselhaus} G. Dresselhaus, Phys. Rev. {\bf 106}, 76
(1957); 

R. J. Elliott, Phys. Rev. {\bf 108}, 1384 (1957).



\bibitem{elliott2} R. J. Elliott and R. Loudon, J. Phys. Chem. Solids.{\bf
8}, 382 (1959); 

R. J. Elliott and R. Loudon,
J. Phys. Chem. Solids.{\bf 15}, 196 (1960).


\bibitem{reed} M. Reed and B. Simon,  {\it Fourier Analysis, 
Self-Adjointness II}
  ( New York :Academic, 1975 ).



\bibitem{kumar1} B. Basu-Mallick, Pijush K. Ghosh and Kumar S. Gupta,
Nucl. Phys. {\bf B659}, 437 (2003).

\bibitem{kumar2} B. Basu-Mallick, Pijush K. Ghosh and Kumar S. Gupta,
Phys. Lett. {\bf A311},  87 (2003).


\bibitem{kumar3} Kumar S. Gupta,  Mod. Phys. Lett. {\bf A18}, 2355 (2003).

\bibitem{biru} B. Basu-Mallick and Kumar S. Gupta, Phys. Lett. {\bf A292}, 36
(2001).


\bibitem{stjep} S. Meljanac, A. Samsarov, B. Basu-Mallick and Kumar S. Gupta,
Eur.  Phys. J. {\bf C49}, 875 (2007).


\bibitem{giri} P. R. Giri, K. S. Gupta, S. Meljanac and A. Samsarov, hep-th/0703121.

\bibitem{bh} D. Birmingham, Kumar S. Gupta and Siddhartha Sen,
Phys. Lett. {\bf B505}, 191 (2001); 

Kumar S. Gupta and Siddhartha
Sen, Phys. Lett. {\bf B526}, 121 (2002).

\bibitem{lamb} W. Lamb, Phys. Rev. {\bf 85}, 259 (1952).

\bibitem{schiff} L. I. Schiff and H. Snyder, Phys. Rev. {\bf 55}, 59 (1939).


\bibitem{dunford} N. Dunford and J. T. Schwartz, {\it Linear
  Operators, Spectral Theory, Self Adjoint Operators in Hilbert Space, Part 2}
  (Wiley-Interscience, 1988).

\bibitem{abr} M. Abromowitz, I. A. Stegun, {\it Handbook of Mathematical
Functions} (Dover, New York, 1970).

\bibitem{feher} L. Feher, I. Tsutsui and T. Fulop, Nucl. Phys. {\bf
B715}, 713 (2005).



\end{thebibliography}
\end{document}